\documentclass[review,authoryear]{elsarticle}

\usepackage{amsthm,amsmath,amsfonts,amssymb}
\usepackage{natbib}

\usepackage[latin1]{inputenc}
\usepackage[english]{babel} % hyphenation patterns for American English
\usepackage{subfigure}
\usepackage{setspace}
\usepackage{tabularx} % para larguras das colunas das tabelas
\usepackage{ctable} % for \toprule, \midrule and \bottomrule
\usepackage{enumerate}
\usepackage{epsfig}
\usepackage{graphics}
\usepackage{graphicx}
\usepackage{color}
\usepackage{booktabs}
\usepackage{longtable}
\usepackage{url}
\usepackage{dsfont}

\usepackage{dsfont}
\usepackage{placeins}
\usepackage{mathrsfs}

\usepackage[short,12hr]{datetime}
\usepackage{fullpage}
\usepackage{srcltx}
\usepackage{flushend}
\usepackage{mathptmx}

\usepackage[normalem]{ulem}

\newcommand{\F}{\mathscr{F}}
\newcommand{\E}{\mathds{E}}
\newcommand{\Var}{\operatorname{Var}}
\newcommand{\diag}{\operatorname{diag}}
\newcommand{\R}{\mathds{R}} % mathds é uma fonte mais bonita para R, Z e C
\newcommand{\Z}{\mathds{Z}}
\newcommand{\B}{\texttt{B}}
\newcommand{\bs}[1]{\boldsymbol{#1}}

\usepackage{lineno}

\allowdisplaybreaks[4] % para evitar espaços entre fórmulas

\definecolor{ggreen}{RGB}{200,0,200}

\theoremstyle{plain}
\newtheorem{thm}{Theorem}[section]

\journal{Journal of Hydrology}

\newcommand*\patchAmsMathEnvironmentForLineno[1]{%
  \expandafter\let\csname old#1\expandafter\endcsname\csname #1\endcsname
  \expandafter\let\csname oldend#1\expandafter\endcsname\csname end#1\endcsname
  \renewenvironment{#1}%
     {\linenomath\csname old#1\endcsname}%
     {\csname oldend#1\endcsname\endlinenomath}}%
\newcommand*\patchBothAmsMathEnvironmentsForLineno[1]{%
  \patchAmsMathEnvironmentForLineno{#1}%
  \patchAmsMathEnvironmentForLineno{#1*}}%
\AtBeginDocument{%
\patchBothAmsMathEnvironmentsForLineno{equation}%
\patchBothAmsMathEnvironmentsForLineno{align}%
\patchBothAmsMathEnvironmentsForLineno{flalign}%
\patchBothAmsMathEnvironmentsForLineno{alignat}%
\patchBothAmsMathEnvironmentsForLineno{gather}%
\patchBothAmsMathEnvironmentsForLineno{multline}%
}

\begin{document}

%\linenumbers
%\begin{linenumbers}
%\begin{linenomath}

%\doublespacing

\begin{frontmatter}

\title{
Kumaraswamy autoregressive moving average models for double bounded environmental data
}

 \author[label1]{F\'abio~Mariano~Bayer\corref{cor1}}
 \ead{bayer@ufsm.br}
 \address[label1]{Departamento de Estat\'istica and LACESM, Universidade Federal de Santa Maria, Santa Maria, Brazil}
 \cortext[cor1]{Principal corresponding author}

 \author[label2]{D\'ebora~Missio~Bayer}
 \ead{debora.bayer@ufsm.br}
 \address[label2]{Departamento de Engenharia Sanit\'aria e Ambiental, Universidade Federal de Santa Maria, Santa Maria, Brazil}

 \author[label3]{Guilherme~Pumi}
 \ead{guilherme.pumi@ufrgs.br }
 \address[label3]{Departamento de Estat\'istica, Universidade Federal do Rio Grande do Sul, Porto Alegre, Brazil}

\begin{abstract}
In this paper we introduce the Kumaraswamy autoregressive moving average models (KARMA), which is a dynamic class of models for time series taking values in the double bounded interval $(a,b)$ following the Kumaraswamy distribution. The Kumaraswamy family of distribution is widely applied in many areas, especially hydrology and related fields. Classical examples are time series representing rates and proportions observed over time. In the proposed KARMA model, the median is modeled by a dynamic structure  containing  autoregressive and moving average terms, time-varying regressors, unknown parameters and a link function. We introduce the new class of models and discuss conditional maximum likelihood estimation, hypothesis testing inference, diagnostic analysis and forecasting. In particular, we provide closed-form expressions for the conditional score vector and conditional Fisher information matrix. An application to environmental real data is presented and discussed.

\end{abstract}

\begin{keyword}
ARMA
\sep
Double bounded data
\sep
Dynamic model
\sep
Forecasts
\sep
Kumaraswamy distribution
\end{keyword}

\end{frontmatter}

\section{Introduction}\label{S:introduction}

The Kumaraswamy family of distribution was introduced in \cite{Kumaraswamy1980} for modeling double bounded random processes with hydrological applications.
It is very flexible being able to approximate several types of distributions and its density can present many shapes, such as unimodal, uniantimodal, increasing, decreasing or constant.
The Kumaraswamy distribution has been applied to a wide variety of problems, especially in hydrology \citep{Nadarajah2008}.

We say that a continuous random variable $\tilde Y$ follows a Kumaraswamy distribution with shape parameters $\varphi >0$ and $\delta>0$ and support in $(a,b)$ if its probability density function is given by \citep{mitnik2013new}:
\begin{align*}
f(\tilde{y};\varphi,\delta)=\frac{\varphi \delta}{b-a}
\left( \frac{\tilde{y}-a}{b-a} \right) ^{\varphi-1}
\left[ 1- \left( \frac{\tilde{y}-a}{b-a}\right)^\varphi \right]^{\delta-1},
\end{align*}
for $a < \tilde{y} < b$. In this case we denote $\tilde Y\sim K(\delta,\varphi,a,b)$. The cumulative distribution and quantile functions are, respectively, given by:
\begin{align*}
F(\tilde{y})= 1 -
\left[1- \left(\frac{\tilde{y}-a}{b-a} \right)^\varphi \right]^\delta\quad \mbox{ and } \quad
F^{-1}(u)=a+(b-a) \left[ 1-(1-u)^\frac{1}{\delta} \right]^\frac{1}{\varphi}
,
\quad
0<u<1.
\end{align*}
The mean and variance of $\tilde{Y}$ are given by
\begin{align}\label{exp}
\E(\tilde{Y})= a + (b-a) \delta \B \left(1+\frac{1}{\varphi},\delta \right)
\end{align}
and
\begin{align}\label{e:var-yab}
\Var(\tilde{Y})= (b-a)^2
\left\lbrace
\delta \B \left( 1+ \frac{2}{\varphi},\delta \right)-
\left[ \delta \B \left (1+ \frac{1}{\varphi},\delta \right) \right]^2
\right\rbrace
,
\end{align}
where $\B(\cdot,\cdot)$ is the Beta function \citep{Gupta2004}.
Some properties of the Kumaraswamy distribution can be found in \cite{Jones2009}, \cite{Gupta2004}, and \cite{mitnik2013new}.
It is usually applied as a more tractable alternative to the beta distribution \citep{Jones2009, Nadarajah2008,mitnik2013kumaraswamy}.

The beta distribution has been widely applied to model double bounded variates.
The beta regression model \citep{Ferrari2004},
for example,
which resembles the generalized linear models (GLM) \citep{McCullagh1989},  has been generalized, improved and applied in several works \citep{Cribari2010, Simas2010, Ospina2012, Bayer2013, Souza2015}.
Beta related models for time series have also received attention in the literature.
See for instance
\cite{Rocha2009},
\cite{daSilva2011},
\cite{Guolo2014},
\cite{Palm2017}
 and references therein.

The flexibility of the beta distribution encourages its empirical use in a wide range of applications \citep{Lemonte2013, Jones2009, Nadarajah2008}.
However,
the beta distribution does not satisfactorily fit hydrological process such as daily rainfall, daily stream flow, etc \citep{Kumaraswamy1976, Kumaraswamy1980, Lemonte2013}.
On the other hand, in hydrology and related areas, the Kumaraswamy distribution is deemed as a better alternative to the beta distribution \citep{Nadarajah2008, Lemonte2013},
so that several works applying the Kumaraswamy distribution can be found \citep{Nadarajah2008}. This is also true in the Engineering literature, as, for instance, in \cite{Sundar1989}, \cite{Fletcher1996}, \cite{seifi2000}, \cite{ponnambalam2001}, \cite{ganji2006}, and  \cite{Koutsoyiannis1989}.

Despite its importance and wide range of applications in hydrology, the Kumaraswamy distribution is still a stranger for statisticians.
In fact,
the lack of tractable expressions for the
mean and variance,
given by \eqref{exp} and \eqref{e:var-yab},
respectively,
has hindered its utilization for modeling purposes  \citep{Lemonte2013, mitnik2013kumaraswamy}.
An alternative to circumvent this problem
is to consider
a median-based re-parameterization aiming to facilitate its use in regression-based models \citep{mitnik2013kumaraswamy}.
For the Kumaraswamy distribution, the median has the following simple expression:
\begin{align*}%\label{med}
md(\tilde{Y})=
\tilde{\mu}=
a+(b-a)\left(1-0.5^\frac{1}{\delta}\right)^\frac{1}{\varphi},
\end{align*}
where $\left(1-0.5^\frac{1}{\delta}\right)^\frac{1}{\varphi}=\mu$ is the median of the rescaled variable
$Y=\frac{\tilde{Y}-a}{b-a}\in (0,1)$.

Most time series appearing in natural sciences,
including hydrology, climatology and environmental applications
consist of observations that are serially dependent overtime \citep{Salas1997, Machiwal2012, Lohani2012, Valipour2013}.
Most conventional time series models are based on Gaussianity assumptions~\citep{Chuang2007}.
One classical example is the class of autoregressive integrated moving average models  (ARIMA)~\citep{Box2008, Brockwell1991}.
However, it has been recognized that the Gaussian assumption is too restrictive for many applications~\citep{Tiku2000},
specially in hydrology.
Indeed, as previously discussed, several double bounded hydrologic data
can be accurately modeled by the Kumaraswamy distribution.
Despite of this, to the best of our knowledge, a
specific time series model to
serially dependent Kumaraswamy variables has never been considered in the literature.
Thus, in this work our goal is to introduce and study a dynamic time series model for Kumaraswamy distributed random variables.
In order to define the proposed model,
we shall follow similar construction as the generalized autoregressive and moving average model (GARMA) \citep{Benjamin2003} and the beta autoregressive and moving average model ($\beta$ARMA) \citep{Rocha2009}, but we shall employ a parametrization for the Kumaraswamy distribution in terms of its median.

The paper is organized as follows. In Section \ref{s:modelo} we introduce the proposed model. In Section \ref{s:estimation} we present a complete conditional maximum likelihood theory for KARMA models, including closed forms for the conditional score vector and Fisher information matrix and the related asymptotic theory. The construction of confidence intervals and hypothesis testing is also discussed. On Section \ref{s:diag} we present several topics regarding model diagnostics and forecasting. In Section \ref{s:simulation} we present a Monte Carlo simulation study to assess the finite sample performance of the conditional maximum likelihood approach developed. An application of KARMA models to relative humidity data is presented in Section \ref{s:application}. Section \ref{s:conclusion} closes the article. For ease of presentation, some technical results are deferred to the Appendix.

\section{The proposed model}\label{s:modelo}

\begin{figure}
\begin{center}
\subfigure[$\mu=0.25$ and values of $\varphi$ indicated in the graph]
{\label{f:kuma-var-phi}\includegraphics[width=0.4\textwidth]{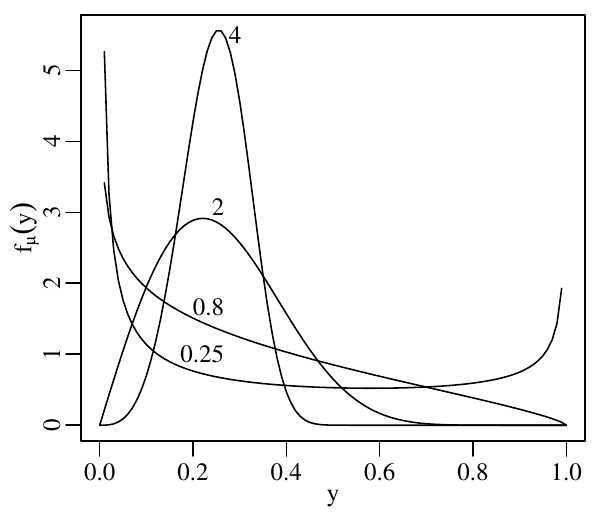}}
\subfigure[$\varphi=2$ and values of $\mu$ indicated in the graph]
{\label{f:kuma-var-mu}\includegraphics[width=0.4\textwidth] {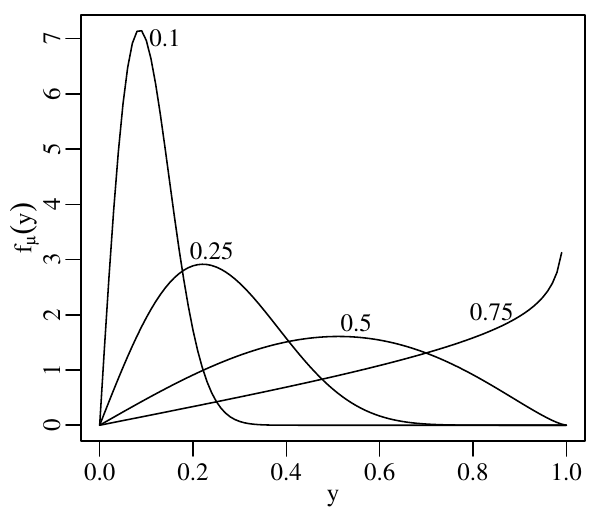}}
\caption{ Kumaraswamy density functions for several values of parameters $\mu$ and $\varphi$ (with $a=0$ and $b=1$).}\label{f:density}
\end{center}
\end{figure}

In order to define the proposed model, we shall introduce an autoregressive moving average (ARMA) time series structure to accommodate the presence of serial correlation in the conditional median of the Kumaraswamy distribution. For this reason, we shall call the proposed model as KARMA model. We shall employ a similar parameterization as in \cite{mitnik2013kumaraswamy}, where
$ \mu=(1-0.5^{\frac{1}{\delta}})^{\frac{1}{\varphi}}$, i.e., $ \delta=\frac{\log(0.5)}{\log(1-\mu^\varphi)}$.
We shall write $\tilde{Y}\sim K(\tilde{\mu},\varphi,a,b)$. Furthermore, $Y=(\tilde{Y}-a)/(b-a)$ follows a $K(\mu,\varphi,0,1)$, where $\mu$ is given by $\mu=(\tilde{\mu}-a)/(b-a)$. For simplicity, in this case we shall write $Y\sim K(\mu,\varphi)$. From this point on we shall only consider the Kumaraswamy distribution parameterized in terms of its median (see Figure~\ref{f:density}).

Now let $\{\tilde{Y}_t\}_{t\in \Z}$ be a stochastic process for which, $\tilde{Y}_t\in(a,b)$ with probability 1, for all $t\in\Z$ and fixed $a,b\in\R$ with $a<b$ and let $\F _{t}=\sigma\{\tilde{Y}_t,\tilde{Y}_{t-1},\dots\}$ denote the sigma-field generated by the information observed up to time $t\in\Z$. Assume that, conditionally to the previous information set $\F _{t-1}$,  $\tilde{Y}_t$ is distributed according to $K(\tilde{\mu}_t,\varphi,a,b)$ and let $Y_t=(\tilde{Y}_t-a)/(b-a)$, $t\in\Z$. It follows that $Y_t|\F _{t-1}\sim K(\mu_t,\varphi)$. The conditional density of $\tilde{Y}_t$ given $\F_{t-1}$ is
\begin{align}\label{model_dens}
f_{\mu_t}(\tilde{y}_t\mid \F _{t-1})
=\left(\frac{1}{b-a}\right)\frac{\varphi\log (0.5)}{\log(1-{\mu}_t^\varphi)} {y}_t^{\varphi-1} (1-{y}_t^\varphi)^{\frac{\log (0.5)}{\log(1-{\mu}_t^\varphi)} -1}, \quad 0<\mu_t<1, \;\;  \varphi>0,
\end{align}
for $0<y_t<1$, where ${y}_t=\frac{\tilde{y}_t-a}{b-a}$, ${\mu}_t = \frac{\tilde{\mu}_t-a}{b-a}$.
This particular form of the density is very appealing since it allows modeling without any transformation, as it is commonly done in literature \citep[see, for instance,][]{Rocha2009}, but dealing with the simpler distribution of $Y_t$.

The cumulative distribution and quantile functions, are given respectively by:
\begin{align*}
F_{\mu_t}(\tilde{y}_t\mid\F _{t-1})
&=
1-(1-{y}_t^\varphi)^{\frac{\log (0.5)}{\log(1-{\mu}_t^\varphi)} },\\
F_{\mu_t}^{-1}(u\mid\F _{t-1})
&=
a+(b-a)\left[
1-(1-u)^\frac{\log(1-{\mu}_t^\varphi)}{\log (0.5)}
\right]
^\frac{1}{\varphi}.
\end{align*}

The conditional mean and variance of $\tilde{Y}_t$, in terms of $\mu_t$ and $\varphi$,
are given respectively by:
\begin{align}
\E(\tilde{Y}_t \mid \F _{t-1} )&=a+(b-a)\frac{\log (0.5)}{\log(1-{\mu}_t^\varphi)} \B \left(1+\frac{1}{\varphi},\frac{\log (0.5)}{\log(1-{\mu}_t^\varphi)} \right), \nonumber \\
\Var(\tilde{Y}_t \mid \F _{t-1})&=\frac{1}{b-a}\bigg\{
\frac{\log (0.5)}{\log(1-{\mu}_t^\varphi)} \B \left( 1+ \frac{2}{\varphi},\frac{\log (0.5)}{\log(1-{\mu}_t^\varphi)} \right) % \nonumber \\
-\left[ \frac{\log (0.5)}{\log(1-{\mu}_t^\varphi)} \B \left (1+ \frac{1}{\varphi},\frac{\log (0.5)}{\log(1-{\mu}_t^\varphi)} \right) \right]^2\bigg\}.\label{e:var}
\end{align}

Let $g:\R\rightarrow(0,1)$ be a continuously twice differentiable monotone link function for which the inverse $g^{-1}:(0,1)\rightarrow\R$ exists and is twice continuously differentiable as well.
We propose the following specification for the conditional median $\mu_t$:
\begin{align}\label{model}
\eta_t=g(\mu_t)= \alpha+x_t^\top \beta + \sum_{i=1}^p \phi_i \big[ g(y_{t-i})-x^\top_{t-i} \beta\big] + \sum_{j=1}^q \theta_j r_{t-j},
\end{align}
where $\eta_t$ is the linear predictor, $x_t$ is the $r$-dimensional vector containing the covariates at time $t$, $\beta=(\beta_1,\dots,\beta_r)^\top$ is the $r$-dimensional vector of parameters related to the covariates, while $\phi=(\phi_1,\dots,\phi_p)^\top$ and $\theta=(\theta_1,\dots,\theta_q)^\top$ are the AR and MA coefficients, respectively. As usual, we shall assume that the AR and MA characteristic polynomials do not have common roots and the AR coefficients are such that the related characteristic polynomial does not have unit roots. Invertibility and causality conditions for the ARMA component are not needed and, thus,  not required.
For more details on ARMA modeling, we refer the reader to \cite{Brockwell1991}.
Observe that, since $\mu_t\in(0,1)$, all the traditional link functions such as logit, probit, loglog, etc, can be applied to the model. Results related to $\tilde{\mu}_t$ and  $\tilde{y}_t$, such as predicted values and confidence intervals, can be easily obtained through $\tilde{\mu}_t=\mu_t(b-a)+a$ and $\tilde{y}_t=y_t(b-a)+a$.

The proposed KARMA$(p,q)$ model is given by specification \eqref{model_dens} and \eqref{model}. We observe that the dynamic part of the model \eqref{model} is the same as in \cite{Rocha2009}.
However, the random component of the model \eqref{model_dens} is completely different and
it is parametrized in terms of the median.
Regression methods based on the median are known to be robust against atypical observations in the response \citep{John2015, Lemonte2016}.
It is also known that, compared to mean based models, median based ones present a better performance when the population distribution is asymmetric \citep{Lemonte2016}. Since the specification \eqref{model} is based on the median, the proposed KARMA model inherits these attractive properties.

\section{Conditional likelihood inference}\label{s:estimation}

Parameter estimation can be carried out by conditional maximum likelihood. The boundary parameters $a$ and $b$ are assumed to be  either known (as it is often the case for rates and proportions) or previously consistently estimated. Let $\tilde y_1,\dots,\tilde y_n$ be a sample from a KARMA$(p,q)$ model under specification \eqref{model_dens} and \eqref{model},
where $x_t$ denote the $r$-dimensional vector of covariates for $y_t$, assumed to be non-stochastic and let $\gamma=(\alpha, \beta^\top, \phi^\top, \theta^\top, \varphi)^\top$ be the $(p+q+r+2)$-dimensional parameter vector. The conditional maximum likelihood estimators (CMLE) are obtained upon maximizing the logarithm of the conditional likelihood function. We observe that the log-likelihood function for $\gamma$, conditionally on $\F_{t-1}$, is null for the first $m=\max(p,q)$ values of $t$, and hence we have
\begin{align}\label{E:loglik}
\ell=\ell(\gamma; \tilde{y}_t)=\sum\limits_{t=m+1}^{n} \log \big(f(\tilde{y}_t\mid\F _{t-1})\big)= \sum\limits_{t=m+1}^{n}\ell_t(\mu_t,\varphi),
\end{align}
where
\[\ell_t(\mu_t,\varphi)=\log (\varphi)\!-\!\log(b\!-\!a)+\log\left(\!\frac{\log (0.5)}{\log(1\!-\!{\mu}_t^\varphi)}\!\right) \!+\!(\varphi\!-\!1)\log ({y}_t)\!+\!\left(\!\frac{\log (0.5)}{\log(1\!-\!{\mu}_t^\varphi)} \!-\!1\!\! \right)\log(1\!-\!{y}_t^\varphi).\]

\subsection{Conditional score vector}\label{s:score}

Recall that $\eta_t=g(\mu_t$). By differentiating the conditional log-likelihood function given in \eqref{E:loglik}, with respect to the $j$th element of the parameter vector $\gamma$, $\gamma_j \neq \varphi$, for $j=1,\dots,(p+q+r+1)$, the chain rule yields
\begin{align*}
 \frac{\partial \ell}{\partial \gamma_j} &=
 \sum\limits_{t=m+1}^{n} \frac{\partial \ell_t(\mu_t,\varphi)}{\partial \mu_t}
 \frac{d \mu_t}{d \eta_t} \frac{\partial \eta_t}{\partial \gamma_j}.
\end{align*}
Now, since $\frac{d \mu_t}{d \eta_t} = \frac{1}{g'(\mu_t)}$,
\begin{align}\label{e:dldmu}
\frac{\partial \ell_t(\mu_t,\varphi)}{\partial \mu_t} =
 \varphi\frac{{\mu}_t^{\varphi-1}}{(1-{\mu}_t^\varphi)\log(1-{\mu}_t^\varphi)}
\left(\delta_t\log(1-{y}_t^\varphi)+1\right)=\varphi c_t,
\end{align}
where, for simplicity, we wrote
\begin{align}\label{e:ct}
c_t=
%c_t(\mu_t,\varphi)=
\frac{{\mu}_t^{\varphi-1}}{(1-{\mu}_t^\varphi)\log(1-{\mu}_t^\varphi)} \left( \delta_t\log(1-{y}_t^\varphi)+1 \right),
\quad \mbox{and} \quad
\delta_t=
%\delta_t(\mu_t,\varphi)=
\frac{\log (0.5)}{\log(1-{\mu}_t^\varphi)}.
\end{align}
We can write
\begin{align*}
 \frac{\partial \ell}{\partial \gamma_j} &=  \sum\limits_{t=m+1}^{n} \frac{\varphi {\mu}_t^{\varphi-1}}{g'(\mu_t)(1-{\mu}_t^\varphi)\log(1-{\mu}_t^\varphi)}
 \left( \delta_t\log(1-{y}_t^\varphi)+1 \right) \frac{\partial \eta_t}{\partial \gamma_j}=\sum\limits_{t=m+1}^{n}\varphi\frac{c_t}{g'(\mu_t)}\frac{\partial \eta_t}{\partial \gamma_j}.
\end{align*}
Observe that the task of computing the score vector greatly simplifies to obtain $\frac{\partial \eta_t}{\partial \gamma_j}$ for each coordinate $\gamma_j$ of $\gamma$. For the derivative of $\ell$ with respect to $\alpha$, let $r_{t}=g(y_t)-g(\mu_t)$ be the
error term, so that
\begin{align*}
\frac{\partial \eta_t}{\partial \alpha}=1 + \sum_{j=1}^q \theta_j \frac{\partial r_{t-j}}{\partial \alpha}=1 - \sum_{j=1}^q \theta_j \frac{\partial \eta_{t-j}}{\partial \alpha}.
\end{align*}
For the derivative of $\ell$ with respect to $\beta$, for $l=1,\dots,r$, we have
\begin{align*}
\frac{\partial \eta_t}{\partial \beta_l}= x_{tl}-\sum_{i=1}^{p} \phi_ix_{(t-i)l}
- \sum_{j=1}^q \theta_j \frac{\partial \eta_{t-j}}{\partial \beta_l},
\end{align*}
where $x_{tl}$ is the $l$th element of $x_t$.  For the derivative of $\ell$ with respect to $\phi$, for $i=1,\dots, p$,
\begin{align*}
\frac{\partial \eta_t}{\partial \phi_i}= g({y}_{t-1})-x^\top_{t-1}\beta
- \sum_{j=1}^q \theta_j \frac{\partial \eta_{t-j}}{\partial \phi_i}.
\end{align*}
For the derivative of $\ell$ with respect to $\theta$,  for $j=1,\dots, q$, we have
\begin{align*}
\frac{\partial \eta_t}{\partial \theta_j}=r_{t-j}
- \sum_{i=1}^q \theta_i \frac{\partial \eta_{t-i}}{\partial \theta_j}.
\end{align*}
Finally, for the derivative of $\ell$ with respect to $\varphi$, direct differentiation of \eqref{E:loglik} is easier to compute, yielding
\begin{align*}
\frac{\partial \ell_t}{\partial \varphi}=\frac{1}{\varphi}+\log(y_t)+c_t \mu_t \log(\mu_t)-(\delta_t-1)\frac{y_t^{\varphi}\log(y_t)}{(1-y_t^{\varphi})}.
\end{align*}

In matrix form,
the score vector can be written as $U(\gamma)= \left( U_{\alpha}(\gamma), U_{\beta}(\gamma)^\top, U_{\phi}(\gamma)^\top, U_{\theta}(\gamma)^\top, U_{\varphi}(\gamma) \right)^\top$, where
\begin{align*}
U_{\alpha}(\gamma)&= v^\top T c, \qquad U_{\beta}(\gamma)= M^\top T c, \qquad U_{\phi}(\gamma)=  P^\top T c,\qquad U_{\theta}(\gamma)= R^\top T c,\\
\mbox{and }\quad U_{\varphi}(\gamma)&= \frac{n-m}{\varphi}+\sum_{t=m+1}^n \left\lbrace \log(y_t)+c_t \mu_t \log(\mu_t)-(\delta_t-1)\frac{y_t^{\varphi}\log(y_t)}{(1-y_t^{\varphi})}\right\rbrace,
\end{align*}
with $T=\diag\left\lbrace 1/g'\left( {\mu}_{m+1}\right),\dots, 1/g'\left( {\mu}_{n}\right) \right\rbrace$, $v = \left(\frac{\partial \eta_{m+1}}{\partial \alpha}, \dots, \frac{\partial \eta_{n}}{\partial \alpha} \right)^\top$,
$c=(\varphi c_{m+1},\dots,\varphi c_n)^\top$
and $M$, $P$, $R$ be the matrices with dimension $(n-m)\times r$, $(n-m)\times p$ and $(m-n)\times q$, respectively, for which the $(i,j)$th elements are given by
\[M_{i,j}=\frac{\partial \eta_{i+m}}{\partial \beta_j},\quad P_{i,j}=\frac{\partial \eta_{i+m}}{\partial \phi_j}, \quad \mbox{ and } \quad R_{i,j}=\frac{\partial \eta_{i+m}}{\partial \theta_j}.\]

The conditional maximum likelihood estimator of $\gamma$ if it exists, it is obtained as a solution of the system $U(\gamma)=\mathbf{0}$, where $\mathbf{0}$ is the null vector in $\R^{p+q+r+2}$.
There is no closed for for the solution of such a system. Conditional maximum likelihood estimates are, thus, obtained by numerically maximizing the log-likelihood function using a Newton or quasi-Newton nonlinear optimization algorithm; see, e.g., \cite{nocedal1999}. In what follows, we shall use the quasi-Newton algorithm known as Broyden-Fletcher-Goldfarb-Shanno (BFGS) method~\citep{press}.
The iterative optimization algorithm requires initialization.
The
starting values of the constant ($\alpha$),
the regressors parameter ($\beta$)
and the autoregressive ($\phi$) parameters
were selected from an ordinary least squares estimate from a linear regression,
where  $Y=(g(y_{m+1}), g(y_{m+2}), \ldots, g(y_{n}))^\top$ are the responses and the covariates matrix is given by
\begin{align*}
X=
\begin{bmatrix}
1& x_{m1} & x_{m2} & \cdots & x_{mr} & g(y_{m}) & g(y_{m-1}) & \cdots & g(y_{m-p+1}) \\
1& x_{(m+1)1} & x_{(m+1)2} & \cdots & x_{(m+1)r} & g(y_{m+1}) & g(y_{m}) & \cdots & g(y_{m-p+2}) \\
\vdots & \vdots & \vdots & \ddots & \vdots & \vdots & \vdots & \ddots & \vdots \\
1& x_{n1} & x_{n2} & \cdots & x_{nr} & g(y_{n-1}) & g(y_{n-2}) & \cdots & g(y_{n-p}) \\
\end{bmatrix}.
\end{align*}
For the parameter $\theta$,
the starting values are set to zero.

\subsection{Conditional information matrix}\label{s:inf}

In this section we derive the conditional Fisher  information matrix. In order to do that we need to compute the expected values of all second order derivatives.
For $\gamma_i \neq \varphi$ and $\gamma_j \neq \varphi$,
with $i,j\in\{1,\dots,p+q+r+1\}$, it can be shown that
\begin{align*}
\frac{\partial^2\ell_t(\mu_t,\varphi)}{\partial \gamma_i \partial \gamma_j} &= \sum_{t=m+1}^{n}\frac{\partial}{\partial \mu_t}
\left( \frac{\partial \ell_t(\mu_t,\varphi)}{\partial \mu_t}\frac{d \mu_t}{d \eta_t} \frac{\partial \eta_t}{\partial \gamma_j}\right)
\frac{d \mu_t}{d \eta_t} \frac{\partial \eta_t}{\partial \gamma_i} \\
&= \sum_{t=m+1}^{n} \left[ \frac{\partial^2 \ell_t(\mu_t,\varphi)}{\partial \mu_t^2}\frac{d \mu_t}{d \eta_t} \frac{\partial \eta_t}{\partial \gamma_j}
+ \frac{\partial \ell_t(\mu_t,\varphi)}{\partial \mu_t}\frac{\partial}{\partial \mu_t}\left(\frac{d \mu_t}{d \eta_t} \frac{\partial \eta_t}{\partial \gamma_j} \right) \right]
\frac{d \mu_t}{d \eta_t} \frac{\partial \eta_t}{\partial \gamma_i}\,.
\end{align*}
From Lemma 1 in the \ref{L:lema1}, it follows that $\E\big(\partial \ell_t(\mu_t,\varphi)/\partial \mu_t \big| \F _{t-1}\big)=0$ and, thus,
\begin{align}\label{expect_incompleta}
\E\left( \left. \frac{\partial^2\ell_t(\mu_t,\varphi)}{\partial \gamma_i \partial \gamma_j}  \right| \F _{t-1} \right)
&= \sum_{t=m+1}^{n} \E\left( \left. \frac{\partial^2 \ell_t(\mu_t,\varphi)}{\partial \mu_t^2} \right| \F _{t-1} \right)
\left(\frac{d \mu_t}{d \eta_t} \right)^2
\frac{\partial \eta_t}{\partial \gamma_i}
\frac{\partial \eta_t}{\partial \gamma_j}.
\end{align}
Let
\[\lambda_k=
%\lambda_k(\mu_t,\varphi)=
\frac{\mu_t^{k\varphi-2}}{(1-\mu_t^\varphi)^k\log(1-\mu_t^\varphi)^k}.\]
Simple calculus yields
\[A_t=
%A(\mu_t,\varphi)=
\frac{\partial}{\partial \mu_t}\bigg(\frac{\mu_t^{\varphi-1}}{(1-\mu_t^\varphi)\log(1-\mu_t^\varphi)}\bigg)
=\varphi\lambda_2\Big[1+\log(1-\mu_t^\varphi)\Big]+(\varphi-1)\lambda_1,\]
so that, by the multiplication rule we obtain
\begin{align}\label{d2ldmu2}
\frac{\partial^2 \ell_t(\mu_t,\varphi)}{\partial \mu_t^2}&=\varphi A_t\left(\frac{\log(0.5)}{\log(1-\mu_t^\varphi)}\log(1-y_t^\varphi)+1\right)+ \frac{\varphi^2\log(0.5)\lambda_2}{\log(1-\mu_t^\varphi)}\log(1-y_t^\varphi)\nonumber\\
&=\varphi A_t+\varphi \delta_t[A_t+\varphi\lambda_2]\log(1-y_t^\varphi).
\end{align}
By Lemma 1 in the \ref{L:lema1} we have  $\E\big[\log(1-Y_t^\varphi)|\F_{t-1}\big]=-1/\delta_t,$
so that
\begin{align*}
\E\bigg( \frac{\partial^2\ell_t(\mu_t,\varphi)}{\partial \mu_t^2}\bigg|\F_{t-1}\bigg)=-\varphi^2\lambda_2=w_t.
\end{align*}
Finally, taking conditional expectation and from Lemma 2 in the \ref{L:lema2} substituting \eqref{expect} into \eqref{d2ldmu2}, from \eqref{expect_incompleta} we obtain
\begin{align}\label{expect_final}
\E \bigg( \frac{\partial^2\ell}{\partial \gamma_i \partial \gamma_j}  \bigg| \F _{t-1} \bigg)%=\nonumber\\
=\sum_{t=m+1}^{n}\frac{w_t}{g'(\mu_t)^2}
\frac{\partial \eta_t}{\partial \gamma_i}
\frac{\partial \eta_t}{\partial \gamma_j}.
\end{align}
From \eqref{expect_final}, the task of obtaining the information matrix simplifies to obtain the derivatives of $\eta_t$ with respect to the parameters which were previously obtained in Section~\ref{s:score}.

Derivatives with respect to $\varphi$, however, are simpler to obtain directly. For the second derivative of $\ell_t$ with respect to $\varphi$, recall that $c_t$ is  given by \eqref{e:ct} and so we have
\begin{align*}
\frac{\partial c_t}{\partial \varphi}
&=\delta_t\mu_t\lambda_2 \log(\mu_t) \log(1-y_t^\varphi)-\delta_t\mu_t\lambda_1\left[\frac{y_t^\varphi\log(y_t)}{1-y_t^\varphi}\right] +c_t\log(\mu_t)\bigg(\frac{\mu_t^{\varphi}}{(1-\mu_t^\phi) \log(1-\mu_t^{\varphi})}+\frac1{1-\mu_t^{\varphi}}\bigg)
\end{align*}
and
\begin{align*}
\frac{\partial}{\partial \varphi}\left[\frac{y_t^\varphi\log(y_t)}{(1-y_t^{\varphi})}\left(\delta_t-1\right)\right]
&=(\delta_t-1)\frac{y_t^\varphi \log(y_t)^2}{(1-y_t^{\varphi})^2}+ \frac{\delta_t\mu_t^\varphi\log(\mu_t)}{(1-\mu_t^{\varphi})\log(1-\mu_t^{\varphi})}\left[ \frac{y_t^\varphi\log(y_t)}{1-y_t^{\varphi}}\right],
\end{align*}
so that
\begin{align}\label{e:d2ldphi2}
\frac{\partial^2\ell_t(\mu_t,\varphi)}{\partial \varphi^2} &=-\frac{1}{\varphi^2}
+\delta_t\mu_t^2\lambda_2\log(\mu_t)^2\log(1-y_t^\varphi)
-2\delta \mu_t^2 \log(\mu_t) \lambda_1 \left(\frac{y_t^\varphi \log(y_t)}{1-y^\varphi_t} \right)\\
& -(\delta_t-1)\frac{y_t^\varphi\log(y_t)^2}{(1-y_t^{\varphi})^2}\nonumber 
 +c_t\mu_t\log(\mu_t)^2\bigg(\lambda_1 \mu_t^2+\frac1{1-\mu_t^{\varphi}}\bigg)
.
\end{align}

Taking conditional expectation in \eqref{e:d2ldphi2} and substituting the results from Lemma 2 in the \ref{L:lema2}, it follows that
\begin{align*}
\E\bigg(\frac{\partial^2\ell }{\partial \varphi^2}\bigg|\F_{t-1}\bigg)= \frac{m-n}{\varphi^2}
-\sum_{t=m+1}^{n}\bigg\lbrace &\mu_t^2 \lambda_2 \log(\mu_t)^2
+ 2 \delta_t \mu_t^2 \log(\mu_t) \lambda_1 \left(\frac{1-\psi(\delta_t +1)-\kappa}{(\delta_t-1)\varphi} \right)\\
&
+\frac{\delta_t \left( \psi(\delta_t)\big[\psi(\delta_t)+2(\kappa-1)\big]-\psi'(\delta_t)+k_0 \right)}{(\delta_t-2)\varphi^2}\bigg\rbrace,
\end{align*}
where $\psi:(0,\infty)\rightarrow\R$ is the digamma function defined as $\psi(z)=\frac{d}{dz}\log\big(\Gamma(z)\big)$, $\psi'(z)=\frac{d}{dz}\psi(z)$ is the trigamma function, $\kappa=0.5772156649\dots$  is the Euler-Mascheroni constant \citep{gradshteyn2007} and $k_0=\pi^2/6+\kappa^2 -2 \kappa$.
As for the derivative with respect to $\gamma_j\neq\varphi$, we have
\begin{align*}
\frac{\partial \ell_t(\mu_t,\varphi)}{\partial \varphi}=\frac{1}{\varphi}+\log(y_t)+c_t \mu_t \log(\mu_t)-(\delta_t-1)\frac{y_t^{\varphi}\log(y_t)}{(1-y_t^{\varphi})},
\end{align*}
\[ \frac{\partial^2\ell_t(\mu_t,\varphi)}{\partial \varphi\partial \gamma_j}=
c_t\frac{\partial \mu_t\log(\mu_t)}{\partial \gamma_j}+\mu_t\log(\mu_t)\frac{\partial c_t}{\partial \gamma_j}
-\frac{y_t^\varphi\log(y_t)}{1-y_t^{\varphi}}\frac{\partial \delta_t}{\partial \gamma_j}.
\]
Recall that $c_t=\frac{1}{\varphi}\frac{\partial \ell_t(\mu_t,\varphi)}{\partial \mu_t}$ so that
\[\frac{\partial c_t}{\partial\gamma_j}=\frac{\partial c_t}{\partial \mu_t}\frac{\partial \mu_t}{\partial \eta_t}\frac{\partial \eta_t}{\partial \gamma_j}=\frac1{\varphi g'(\mu_t)}\frac{\partial^2 \ell_t(\mu_t,\varphi)}{\partial \mu_t^2}\frac{\partial \eta_t}{\partial \gamma_j},\]
\[\frac{\partial \mu_t\log(\mu_t)}{\partial \gamma_j}=\frac{\partial \mu_t\log(\mu_t)}{\partial\mu_t}\frac{\partial\mu_t}{\partial \eta_t}\frac{\partial \eta_t}{\partial\gamma_j}=\bigg(\frac{\log(\mu_t)+1}{g'(\mu_t)}\bigg)\frac{\partial \eta_t}{\partial\gamma_j},\]
and
\[
\frac{\partial \delta_t}{\partial\gamma_j}
=\frac{\partial \delta_t}{\partial \mu_t}\frac{\partial \mu_t}{\partial \eta_t}\frac{\partial \eta_t}{\partial \gamma_j}
%=\frac1{g'(\mu_t)}\frac{\partial \eta_t}{\partial \gamma_j}\frac{\partial \delta_t}{\partial \mu_t}
=
\frac{1}{g'(\mu_t)}\frac{\partial \eta_t}{\partial \gamma_j}\bigg[\frac{\log(0.5)\varphi\mu_t^{\varphi-1}}{(1-\mu_t^\varphi)\log(1-\mu_t^\varphi)^2}\bigg]
=\bigg(\frac{\varphi\mu_t \delta_t\lambda_1}{g'(\mu_t)}\bigg)\frac{\partial \eta_t}{\partial \gamma_j}\,.
\]
Hence
\begin{align*}
\frac{\partial^2\ell_t(\mu_t,\varphi)}{\partial \varphi\partial \gamma_j}=
\bigg\lbrace
\frac{1}{g'(\mu_t)}
\bigg[c_t(\log(\mu_t)+1)+\frac{\mu_t\log(\mu_t)}{\varphi}\frac{\partial^2 \ell_t(\mu_t,\varphi)}{\partial \mu_t^2}-\varphi\mu_t\delta_t\lambda_1\frac{y_t^{\varphi}\log(y_t)}{(1-y_t^{\varphi})}
\bigg]\bigg\rbrace
\frac{\partial \eta_t}{\partial \gamma_j}
\end{align*}
and upon taking conditional expectation it follows that
\begin{align*}
\E\bigg(\frac{\partial^2\ell }{\partial \varphi\partial \gamma_j}\bigg|\F_{t-1}\bigg)
=\sum_{t=m+1}^{n} \frac{d_t}{g'(\mu_t)}
\frac{\partial \eta_t}{\partial \gamma_j},
\end{align*}
where
$d_t= -\varphi\mu_t\log(\mu_t)\lambda_2-\varphi\mu_t\delta_t\lambda_1\left(\frac{1-\psi(\delta_t+1)-\kappa}{(\delta_t-1)\varphi}\right)$.

Let
$L=\mathrm{diag}\left\{
\E\left(
\frac{\partial^2\ell_{m+1} (\mu_{m+1},\varphi)}{\partial \varphi^2}
\right),
\dots,
\E\left(
\frac{\partial^2\ell_{n} (\mu_{n},\varphi)}{\partial \varphi^2}
\right)
\right\}
$,
$W=\mathrm{diag}\{w_{m+1},\dots,w_n\}$, %ok
and
$D = \mathrm{diag}\{d_{m+1},\dots,d_n\}$.
The joint conditional Fisher information matrix for $\gamma$ is
\begin{align}\label{e:fisher}
K= K(\gamma) = \left( \begin{array}{cccccc}
K_{(\alpha,\alpha)} &K_{(\alpha,\beta)} &K_{(\alpha,\phi)} &K_{(\alpha,\theta)} &K_{(\alpha,\varphi)} \\
K_{(\beta,\alpha)} &K_{(\beta,\beta)} &K_{(\beta,\phi)} &K_{(\beta,\theta)} &K_{(\beta,\varphi)} \\
K_{(\phi,\alpha)} &K_{(\phi,\beta)} &K_{(\phi,\phi)} &K_{(\phi,\theta)} &K_{(\phi,\varphi)} \\
K_{(\theta,\alpha)} &K_{(\theta,\beta)} &K_{(\theta,\phi)} &K_{(\theta,\theta)} &K_{(\theta,\varphi)} \\
K_{(\varphi,\alpha)} &K_{(\varphi,\beta)} &K_{(\varphi,\phi)} &K_{(\varphi,\theta)} &K_{(\varphi,\varphi)}
\end{array} \right),
\end{align}
where
$K_{(\alpha,\alpha)} = -v^\top WT^2 v$,
$K_{(\alpha,\beta)} = K_{(\beta,\alpha)}^{\top} = -v^\top WT^2 M$,
$K_{(\alpha,\phi)} = K_{(\phi,\alpha)}^{\top} = -v^\top WT^2 P$,
$K_{(\alpha,\theta)} = K_{(\theta,\alpha)}^{\top} = -v^\top WT^2 R$,
$K_{(\alpha,\varphi)} = K_{(\varphi,\alpha)}^{\top} = -v^\top DT \textbf{1}$,
$K_{(\beta,\beta)} = -M^\top WT^2 M$,
$K_{(\beta,\phi)} = K_{(\phi,\beta)}^{\top} = -M^\top WT^2 P$,
$K_{(\beta,\theta)} = K_{(\theta,\beta)}^{\top} = -M^\top WT^2 R$,
$K_{(\beta,\varphi)} = K_{(\varphi,\beta)}^{\top} = -M^\top DT \textbf{1}$,
$K_{(\phi,\phi)} = -P^\top WT^2 P$,
$K_{(\phi,\theta)} = K_{(\theta,\phi)}^{\top} = -P^\top WT^2 R$,
$K_{(\phi,\varphi)} = K_{(\varphi,\phi)}^{\top} = -P^\top DT \textbf{1}$,
$K_{(\theta,\theta)} = -R^\top WT^2 R$,
$K_{(\theta,\varphi)} = K_{(\varphi,\theta)}^{\top} = -R^\top DT \textbf{1}$,
%precision
$K_{(\varphi,\varphi)} = -\mathrm{tr}(L)$,
$\textbf{1}$ is an $(n-m)\times1$ vector of ones,
and $\mathrm{tr}(\cdot)$ is the trace function.
We note that the conditional Fisher information matrix is not block diagonal, and hence the parameters are not orthogonal \citep{Cox1987}.

The next Theorem establishes the strong consistency and asymptotic normality of the CMLE for the KARMA$(p,q)$ model. In order to guarantee that the asymptotic variance-covariance matrix is positive definite, we shall need some assumptions on the covariates in the model.
Let
$Z_t=\big(1,\bs{x}_{t-1}^\top,\bs{h}(t,1),\dots,\bs{h}(t,p), r_{t-1},r_{t-2},\dots\big)^\top$ denote the design (covariate) matrix related to (5),
where $\bs{h}(t,j)=g(y_{t-j})-P_{g(y_j)}(\bs{x}_1^\top,\dots,\bs{x}_{t-j-1}^\top)$
and $P_{g(y_j)}(\bs{x}_1^\top,\dots,\bs{x}_{t-j-1}^\top)$ denotes the projection of $g(y_j)$ into the space generated by $\bs{x}_1^\top,\dots,\bs{x}_{t-j-1}^\top$. We assume that $\bs{Z_t}$ belongs to a compact set $\Omega$ of the appropriate real space. We assume further that $\sum_{t=1}^n\bs{Z}_t\bs{Z}_t^\top>0$, for sufficiently large $n$, and that, at the true value of $\bs\gamma$, the matrix $K$ is positive definite for the given set of covariates. A detailed discussion can be found in \cite{Fokianos2004} and \cite{Andersen1970}.

\begin{thm}\label{t:an}
Let $\tilde{y}_1,\dots,\tilde{y}_n$ be a sample from a process following \eqref{model_dens} and \eqref{model}, for known $a<b$ and let $\gamma\in\R^{r+p+q+2}$ denote the true parameter vector.
Let $\widehat\gamma$ denote the CMLE based on the given sample.
Then, as $n$ tends to $\infty$,
\[\widehat{\gamma}\stackrel{\mbox{a.s.}}{\longrightarrow}\gamma,
\;\text{ and } \;
\widehat\gamma \stackrel{\mathcal{D}}{\longrightarrow}\mathcal{N}_s(\gamma, K^{-1}),\]
where $s=r+p+q+2$,
$\mathcal{N}_s(\gamma, K^{-1})$ denotes the $s$-variate normal distribution with mean $\gamma$ and variance-covariance matrix $K^{-1}$,
$\stackrel{\mbox{a.s.}}{\longrightarrow}$ denotes almost sure convergence
and
$\stackrel{\mathcal{D}}{\longrightarrow}$ denotes convergence in distribution.
\end{thm}
The proof of Theorem \ref{t:an} is given in the \ref{a:proofteo}.

\subsection{Confidence intervals and hypothesis testing inference}

The results in Theorem~\ref{t:an} allow the construction of asymptotic confidence intervals/regions and test statistics for hypothesis testing. Let $\tilde{y}_1,\dots,\tilde{y}_n$ be a sample from a KARMA$(p,q)$ model,
$\gamma_{i}$ denote the $i$th component of the true parameter vector $\gamma$
and
$K(\widehat{\gamma})^{ij}$ be the $(i,j)$th element of the inverse of the conditional information matrix \eqref{e:fisher} evaluated at $\widehat{\gamma}\in\R^{p+q+r+2}$,
where $\widehat{\gamma}_i$ is the $i$th coordinate of the CMLE $\widehat{\gamma}$ obtained from the sample.
From the results in Theorem \ref{t:an}, we have
\begin{align*}
\frac{\widehat{\gamma}_i - \gamma_i}{\sqrt{K(\widehat{\gamma})^{ii}}} \stackrel{\mathcal{D}}{\longrightarrow} \mathcal{N}(0,1),
\end{align*}
from which asymptotic confidence intervals for the individual model parameters can be constructed by standard methods. More specifically, let $z_{\delta}$ be the $\delta$ standard normal upper quantile. A $100(1-\alpha)\%$, $0 < \alpha < 1/2$, asymptotic confidence interval for $\gamma_i$, $i=1,\dots,(p+q+r+2)$, is
\begin{align*}
\left[\widehat{\gamma}_i
- z_{1-\alpha/2} \sqrt{K(\widehat{\gamma})^{ii}};\widehat{\gamma}
+ z_{1-\alpha/2} \sqrt{K(\widehat{\gamma})^{ii}}\right].
\end{align*}

We can also apply the results in Theorem~\ref{t:an} to derive asymptotic test statistics for hypothesis testing. Let $\gamma_{i}^0$ be a given hypothesized value for the true parameter $\gamma_{i}$. To test
$\mathcal{H}_0:\gamma_{i}=\gamma_{i}^0$
against
$\mathcal{H}_1:\gamma_{i} \neq \gamma_{i}^0$,
we can apply
an asymptotic version for the signed square root of Wald's statistic \citep{Wald1943}, which is given by~\citep{Pawitan2001}
\begin{align*}
Z= \dfrac{\widehat{\gamma}_i-\gamma_{i}^0}{\sqrt{K(\widehat{\gamma})^{ii}}}\,.
\end{align*}
Under $\mathcal{H}_0$, the limiting distribution of $Z$ is standard normal.
Thus,
the test is performed by comparing the calculated $Z$ statistic with the usual quantiles of the standard normal distribution.

From the results in Theorem \ref{t:an}, it is also straightforward to derive versions for
the likelihood ratio~\citep{Person1928}, Rao's score~\citep{Rao1948}, Wald's~\citep{Wald1943}
and the gradient~\citep{Terrell2002} statistics to perform more general hypothesis testing inference. In large samples and under the null hypothesis, such test statistics are (approximately) chi-squared distributed with the same degrees of freedom as their counterparts under independence.

\section{Diagnostic analysis and forecasting}\label{s:diag}

This section introduces some diagnostic measures and forecasting methods.
Diagnostic analysis can be applied to a fitted model to determine whether it fully captures the data dynamics.
A fitted model that passes all diagnostic checks can be used for out-of-sample forecasting.

Information criteria are important tools for automatic model comparison/selection.
Information criterion such as
Akaike's (AIC)~\citep{Akaike1974},
Schwartz's (SIC)~\citep{Schwarz1978},
and Hannan and Quinn's (HQ)~\citep{Hannan1979}
are obtained in the usual fashion from the maximized conditional log-likelihood function.

Residuals are an important measure for determining whether the fitted model provides a good fit to the data \citep{Kedem2002}.
Various types of residuals are currently available in literature for several classes of models \citep{Mauricio2008}.
For the proposed KARMA$(p,q)$,
the standardized (or Person's)
or
deviance
residuals
can be considered.
However,
we suggest the quantile residuals \citep{Dunn1996},
that possess
several advantages over other residuals.
The quantile residuals are defined by
\[r^{(q)}_t=\Phi^{-1}\big(F_{\mu_t}(\tilde y_t|\F_{t-1})\big),\]
where $\Phi^{-1}$ denotes the standard normal quantile function.
The quantile residuals not only can detect lack of fit in regression models but its distribution is also approximately standard normal \citep{Dunn1996, Pereira2017}.
If the model provides a good fit to the data,
the index plot of the quantile residuals should display no noticeable pattern.

When the model is correctly specified the residuals should display white noise behavior, i.e., they should follow a zero mean and constant variance uncorrelated process \citep{Kedem2002}.
A good alternative to test the adequacity of the fitted model is to deploy a Ljung-Box type test \citep{Ljung1978} based on the residual.
More details can be found in \cite{greene2003} and references therein.

Forecasting the conditional median of a KARMA$(p,q)$ model can be done using the theory of time series forecasting for ARMA models \citep{Brockwell1991,Box2008}. Let $h_0$ denote the forecast horizon. We shall assume that the covariate values $x_t$, for $t=n+1,\dots,n+h_0$, are available or can be obtained. For instance, if the covariates are deterministic functions of $t$, as for instance, sines and cosines in harmonic analysis, dummy variables, polynomial trends, etc, they can be determined for values of $t>n$.

The first step is to obtain the estimates $\widehat{\mu}_{m+1},\dots,\widehat{\mu}_n$ for the conditional median $\mu_t$ based on the CMLE $\widehat{\gamma}$. To do that we need to recompose the error term $\{r_t\}_{t=1}^n$, which we will denote by $\widehat{r}_t$. We start by setting $\widehat{r}_t=\E(r_t)$, which usually equals 0, for $t\in\{1,\dots,m\}$. Starting at $t=m+1$, we sequentially set
\[\widehat\mu_t=g^{-1} \bigg( \widehat{\alpha}+ x_t^\top\widehat\beta
+ \sum_{i=1}^{p}\widehat{\phi}_i \big( g(y_{t-i})- x_{t-i}^\top\widehat\beta\big)
+\sum_{j=1}^{q}\widehat{\theta}_j \widehat{r}_{t-j} \bigg)\]
and $\widehat{r}_t=g(y_{t})-g(\widehat\mu_t)$, for $t\in\{m+1,\dots,n\}$. Now, for $h=1,2,\dots,h_0$, the forecasted values of $\mu_{n+h}$ are sequentially  given by
\[\widehat\mu_{n+h}=g^{-1} \bigg( \widehat{\alpha}+ x_{n+h}^\top\widehat\beta
+ \sum_{i=1}^{p}\widehat{\phi}_i \Big([g(y_{n+h-i})]- x_{n+h-i}^\top\widehat\beta\Big)
+\sum_{j=1}^{q}\widehat{\theta}_j \widehat{r}_{n+h-j} \bigg),\]
where $\widehat{r}_t=0$, for $t>n$, and
\begin{align*}
\left[ g(y_{t}) \right]  =\left\{\begin{array}{rc}
g(\widehat{\mu}_{t}),&\textrm{if}\;\;\; t>n,\\
g(y_{t}), &\textrm{if}\;\;\; t\leq n.
\end{array}\right.
%\quad \text{and} \quad
%\widehat r_{t} =\left\{\begin{array}{rc}
%0,&\textrm{if}\;\;\; t>n,\\
%\widehat{r}_t, &\textrm{if}\;\;\; t\leq n,
%\end{array}\right. .
\end{align*}

\section{Numerical evaluation} \label{s:simulation}

In this section we present a Monte Carlo simulation study to assess the finite sample performance of the CMLE for KARMA models developed in Sections \ref{s:score} and \ref{s:inf}.
We simulate 10,000 replications of a KARMA$(2,2)$ model restricted to the interval $(0,1)$ with parameters $\alpha=0.50$, $\phi=(0.50,-0.30)$, $\theta = (0.40,0.15)$, $\varphi=15$
and
a KARMA$(1,1)$ restricted to the $(0,1)$ with parameters $\alpha=-1.00$, $\phi_1=-0.50$, $\theta_1 = 0.25$, $\varphi=10$.
The sample sizes considered are $n\in\{70,100,200,300\}$,
the link function is the logit
and no covariates were included in the simulations.

To generate a size $n$ sample from a KARMA$(p,q)$ process, the following algorithm is useful. The first step is to set $r_t=0$ and $\mu_t=g^{-1}(\alpha)$, for $t=1,\dots,m$. Second step: for $t=m+1$, we obtain $\eta_t$ through \eqref{model}, then we set $\mu_t=g^{-1}(\eta_t)$. Finally, $\tilde{y}_t$ is generated from \eqref{model_dens}, using any adequate method. The so-called inversion method is very easy to apply in this context, that is, we generate $u\sim U(0,1)$ and set $\tilde y_t=a + (b-a)\Big( 1-(1-u)^{ \log(1-\mu_t^\varphi)/\log(0.5)} \Big)^{1/\varphi}$. We iterate the second step for $t=m+1,\dots,n_0+n$, where $n_0>m$ denotes the size of a possible burn in. We used $n_0=2m$ in the simulations. The desired sample is $\tilde{y}_{n_0+1},\dots,\tilde{y}_{n_0+n}$. All routines were written in R language by the authors and are available upon request.

\begin{table}[p]
\caption{Monte Carlo simulation results for the CMLE estimator based on KARMA$(2,2)$ model.
} \label{t:simu1}
\centering
\begin{tabular}{lrrrrrr}
\hline
	& $	\alpha	$ & $	\phi_1	$ & $	\phi_2	$ & $	\theta_1	$ & $	\theta_2	$ & $	 \varphi	$ \\
%\hline
Parameter	& $	0.5000	$ & $	0.5000	$ & $	-0.3000	$ & $	0.4000	$ & $	0.1500	$ & $	 15.0000	$ \\
\hline
	\multicolumn{7}{c}{$n=70$}\\
\hline
Mean	& $	0.4490	$ & $	0.6399	$ & $	-0.3625	$ & $	0.2362	$ & $	0.0766	$ & $	16.0342	 $ \\
%Bias	& $	-0.0510	$ & $	0.1399	$ & $	-0.0625	$ & $	-0.1638	$ & $	-0.0734	$ & $	1.0342	 $ \\
RB (\%)	& $	-10.2065	$ & $	27.9743	$ & $	20.8497	$ & $	-40.9584	$ & $	-48.9391	$ & $	6.8947	$ \\
%SD	& $	0.2141	$ & $	0.4306	$ & $	0.1837	$ & $	0.4850	$ & $	0.3307	$ & $	1.7389	$ \\
%Var	& $	0.0459	$ & $	0.1854	$ & $	0.0337	$ & $	0.2352	$ & $	0.1093	$ & $	3.0239	 $ \\
MSE	& $	0.0485	$ & $	0.2050	$ & $	0.0376	$ & $	0.2621	$ & $	0.1147	$ & $	4.0935	$ \\
%\hline
%	& $	\alpha	$ & $	\phi_1	$ & $	\phi_2	$ & $	\theta_1	$ & $	\theta_2	$ & $	 \varphi	$ \\
%\hline
%	& $	0.5000	$ & $	0.5000	$ & $	-0.3000	$ & $	0.4000	$ & $	0.1500	$ & $	15.0000	$ \\
\hline
\multicolumn{7}{c}{$n=100$}\\
\hline
Mean	& $	0.4605	$ & $	0.6056	$ & $	-0.3440	$ & $	0.2851	$ & $	0.0938	$ & $	15.6546	 $ \\
%Bias	& $	-0.0395	$ & $	0.1056	$ & $	-0.0440	$ & $	-0.1149	$ & $	-0.0562	$ & $	0.6546	 $ \\
RB (\%)	& $	-7.8992	$ & $	21.1131	$ & $	14.6614	$ & $	-28.7169	$ & $	-37.4870	$ & $	 4.3643	$ \\
%SD	& $	0.1883	$ & $	0.3894	$ & $	0.1596	$ & $	0.4150	$ & $	0.2743	$ & $	1.3209	$ \\
%Var	& $	0.0355	$ & $	0.1516	$ & $	0.0255	$ & $	0.1722	$ & $	0.0752	$ & $	1.7447	 $ \\
MSE	& $	0.0370	$ & $	0.1627	$ & $	0.0274	$ & $	0.1854	$ & $	0.0784	$ & $	2.1732	$ \\
%\hline
%	& $	\alpha	$ & $	\phi_1	$ & $	\phi_2	$ & $	\theta_1	$ & $	\theta_2	$ & $	 \varphi	$ \\
%\hline	
%	& $	0.5000	$ & $	0.5000	$ & $	-0.3000	$ & $	0.4000	$ & $	0.1500	$ & $	15.0000	$ \\
\hline
\multicolumn{7}{c}{$n=200$}\\
\hline
Mean	& $	0.4795	$ & $	0.5550	$ & $	-0.3229	$ & $	0.3434	$ & $	0.1208	$ & $	15.3043	 $ \\
%Bias	& $	-0.0205	$ & $	0.0550	$ & $	-0.0229	$ & $	-0.0566	$ & $	-0.0292	$ & $	0.3043	 $ \\
RB (\%)	& $	-4.1071	$ & $	10.9985	$ & $	7.6245	$ & $	-14.1516	$ & $	-19.4393	$ & $	 2.0288	$ \\
%SD	& $	0.1326	$ & $	0.2854	$ & $	0.1150	$ & $	0.2927	$ & $	0.1872	$ & $	0.8784	$ \\
%Var	& $	0.0176	$ & $	0.0815	$ & $	0.0132	$ & $	0.0857	$ & $	0.0351	$ & $	0.7716	 $ \\
MSE	& $	0.0180	$ & $	0.0845	$ & $	0.0137	$ & $	0.0889	$ & $	0.0359	$ & $	0.8642	$ \\
%\hline
%	& $	\alpha	$ & $	\phi_1	$ & $	\phi_2	$ & $	\theta_1	$ & $	\theta_2	$ & $	 \varphi	$ \\
%\hline
%	& $	0.5000	$ & $	0.5000	$ & $	-0.3000	$ & $	0.4000	$ & $	0.1500	$ & $	15.0000	$ \\
\hline
\multicolumn{7}{c}{$n=300$}\\
\hline
Mean	& $	0.4866	$ & $	0.5354	$ & $	-0.3140	$ & $	0.3640	$ & $	0.1315	$ & $	15.1871	 $ \\
%Bias	& $	-0.0134	$ & $	0.0354	$ & $	-0.0140	$ & $	-0.0360	$ & $	-0.0185	$ & $	0.1871	 $ \\
RB (\%)	& $	-2.6889	$ & $	7.0743	$ & $	4.6832	$ & $	-9.0105	$ & $	-12.3411	$ & $	 1.2470	$ \\
%SD	& $	0.1074	$ & $	0.2345	$ & $	0.0957	$ & $	0.2405	$ & $	0.1498	$ & $	0.7099	$ \\
%Var	& $	0.0115	$ & $	0.0550	$ & $	0.0092	$ & $	0.0578	$ & $	0.0224	$ & $	0.5039	 $ \\
MSE	& $	0.0117	$ & $	0.0562	$ & $	0.0094	$ & $	0.0591	$ & $	0.0228	$ & $	0.5389	$ \\
\hline
\end{tabular}
%\end{table}

\vspace{1cm}

\caption{Monte Carlo simulation results for the CMLE estimator based on KARMA$(1,1)$ model.
} \label{t:simu2}
\centering
\begin{tabular}{lrrrr}
\hline
	& $	\alpha	$ & $	\phi_1	$  & $	\theta_1	$  & $	 \varphi	$ \\
%\hline
Parameter	& $	-1.0000	$ & $	-0.5000	$ & $	0.2500	$ & $	 10.0000	$ \\
\hline
	\multicolumn{5}{c}{$n=70$}\\
\hline
Mean	& $	-0.9385	$ & $	-0.4066	$ & $	0.1513	$ & $	10.4474	$ \\
RB (\%)	& $	-6.1467	$ & $	-18.6736	$ & $	-39.4795	$ & $	4.4737	$ \\
MSE	& $	0.0524	$ & $	0.1183	$ & $	0.1509	$ & $	1.3000	$ \\
\hline
\multicolumn{5}{c}{$n=100$}\\
\hline
Mean	& $	-0.9426	$ & $	-0.4129	$ & $	0.1617	$ & $	10.2745	$ \\
RB (\%)	& $	-5.7441	$ & $	-17.4287	$ & $	-35.3277	$ & $	2.7453	$ \\
MSE	& $	0.0438	$ & $	0.0992	$ & $	0.1120	$ & $	0.7484	$ \\
\hline
\multicolumn{5}{c}{$n=200$}\\
\hline
Mean	& $	-0.9748	$ & $	-0.4618	$ & $	0.2097	$ & $	10.1266	$ \\
RB (\%)	& $	-2.5215	$ & $	-7.6398	$ & $	-16.1322	$ & $	1.2664	$ \\
MSE	& $	0.0175	$ & $	0.0396	$ & $	0.0470	$ & $	0.3420	$ \\
\hline
\multicolumn{5}{c}{$n=300$}\\
\hline
Mean	& $	-0.9830	$ & $	-0.4744	$ & $	0.2234	$ & $	10.0934	$ \\
RB (\%)	& $	-1.6986	$ & $	-5.1255	$ & $	-10.6460	$ & $	0.9338	$ \\
MSE	& $	0.0109	$ & $	0.0247	$ & $	0.0299	$ & $	0.2256	$ \\
\hline
\end{tabular}
\end{table}

Tables~\ref{t:simu1} and \ref{t:simu2} present the simulation results.
Performance statistics presented are the mean,
percentage relative bias (RB\%),
and mean square error (MSE).
The percentage relative bias is defined as the ratio between the bias and the true parameter value times 100.
We observe that the overall performance of the CMLE is very good, except, as expected, for the very small sample size $n=70$. The estimates greatly improve from the case $n=70$ as the sample size increases. Overall the  parameter estimator with the smallest relative bias is $\varphi$ while $\theta_1$ and $\theta_2$ are the ones with the greatest relative bias in all situations.
In general, the estimates perform better in the autoregressive estimator than in the part of the moving averages.
Such fact was already discussed by \cite{Ansley1980} in traditional ARMA models, for example.
It was also verified in $\beta$ARMA model in \cite{Palm2017}.
Thus,
simulation studies show that inferences about parameters of moving averages are usually poorer compared to other parameters.
In all situations, the estimates present small MSE.

\section{Application to relative humidity data}\label{s:application}

The relative air humidity (or simply relative humidity, abbreviated RH) is an important meteorological characteristic to public health, irrigation scheduling design, and hydrological studies. Low RH is known to causes health problems, such as allergies, asthma attacks, dehydration, nasal bleeding, among others \citep{Falagas2008,
Zhang2016}. High RH, on the other hand, is also known to cause respiratory problems, besides being responsible for the increase in precipitation which, in excess, can cause serious consequences, for instance, to urban drainage \citep{Silveira2002}.
The vapor pressure, for example, is a function of the RH and it is an important variable in evapotranspiration estimation methods,
such as the Penman-Monteith \citep{Allen1998},
which is one of the most important and accurate method in hydrology
to estimate evapotranspiration \citep{shuttleworth1992, Allen1998}.
It is also widely applied in physical based hydrological simulations
\citep{collischonn2007,arnold2012}.
Given its relevance, understanding and modeling its behavior is of utmost importance, and so is accurate forecasting of the RH. For instance, it helps the State taking preventive measures regarding public health, management of water resources as well as in climate predictions.

\begin{figure}[t]
\begin{center}
\subfigure[Observed data]
{\label{f:rh-serie}\includegraphics[width=0.4\textwidth] {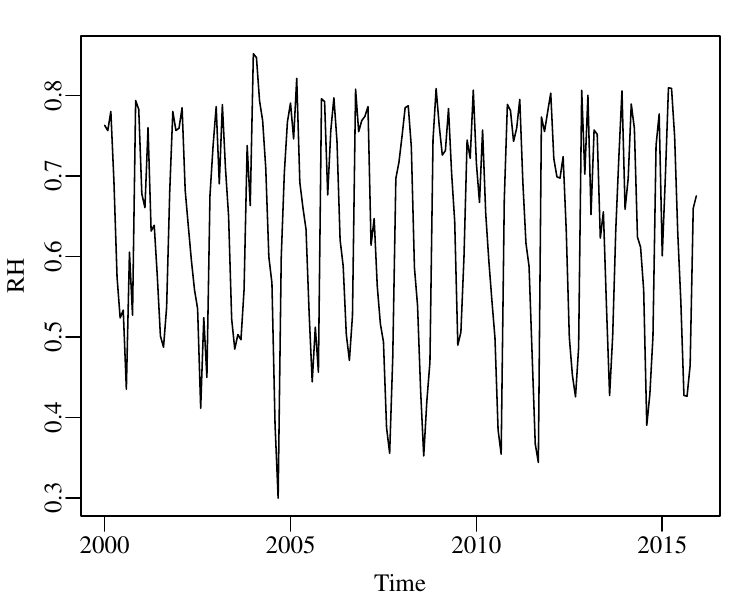}}
\subfigure[Sazonality]
{\label{f:rh-sazonal}\includegraphics[width=0.4\textwidth]{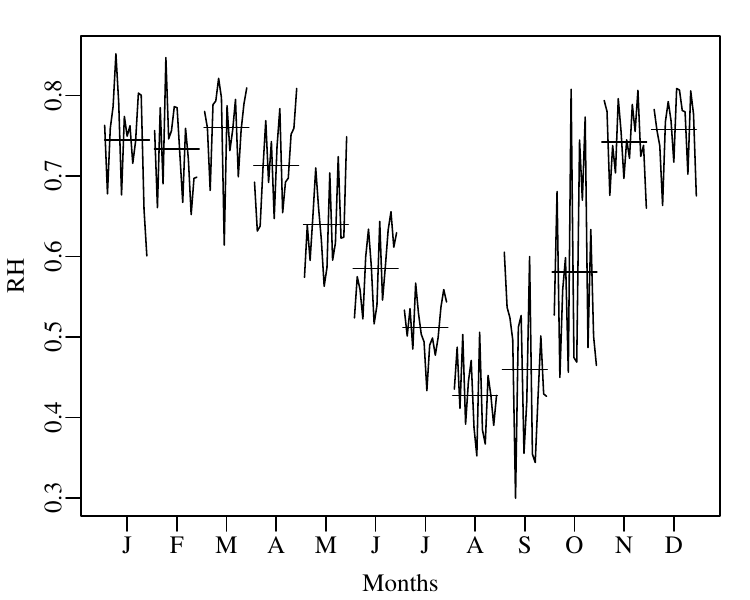}}
\subfigure[ACF]
{\label{f:rh-fac}\includegraphics[width=0.4\textwidth] {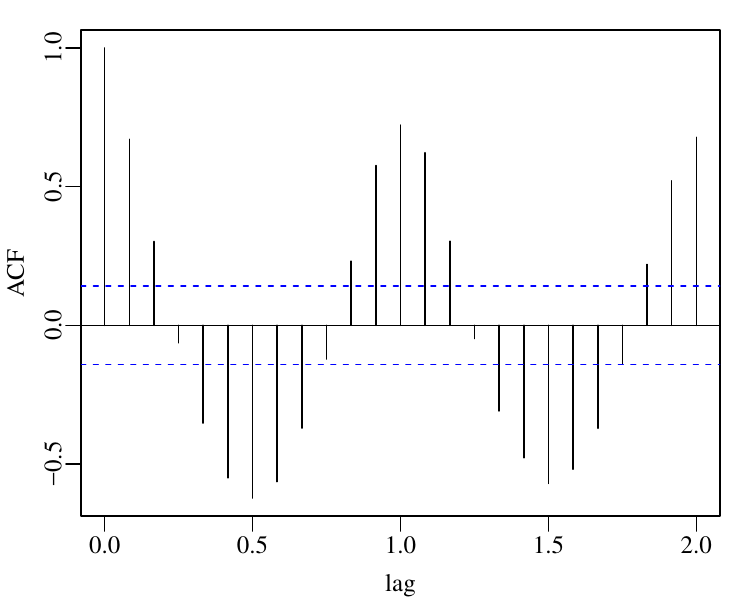}}
\subfigure[PACF]
{\label{f:rh-facp}\includegraphics[width=0.4\textwidth] {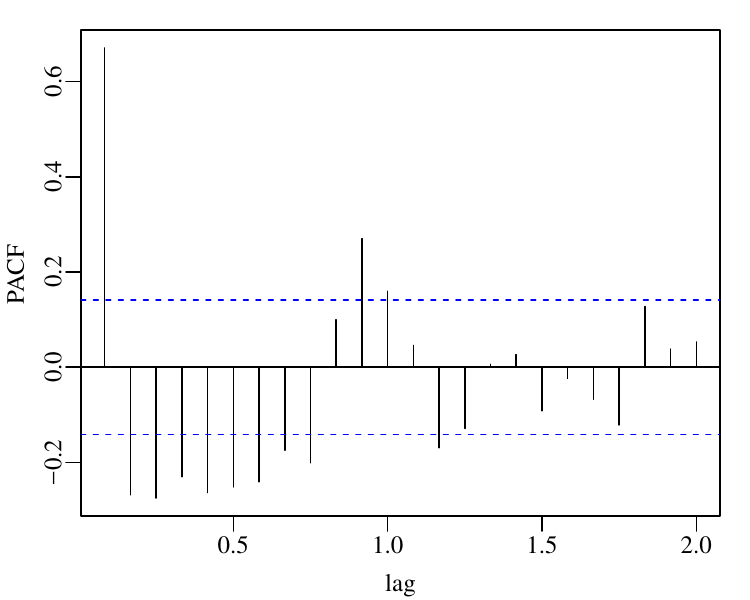}}
\caption{Observed RH time series in Brasilia, Brazil.}\label{f:rh}
\vspace{-0.3cm}
\end{center}
\end{figure}

Relative humidity is an important climate quantity which influences the weather in several ways. The Brazilian capital, Brasilia, is situated in the Center-West Region of Brazil, approximately at latitude 15° 48' south and longitude 48° 55' west.
This particular region in Brazil is plagued by a punishing dry season \citep{Coutinho2002}.
During the dry months, RH often attain dangerously low values, as low as 8\% (measured in July, 7th, 2016 at the Jucelino Kubitshek airport in Brasilia). According to the data from the Instituto Nacional de Meteorologia (INMET - Brazilian National Institute of Meteorological Research), the average annual precipitation in Brasilia is approximately 1500 mm, but from May to September it is specially dry, with a monthly average precipitation of only 13.2 mm (corresponding to about 6\% of the annual average).

The
time series we shall analyze represent the monthly average RH registered in the aforementioned station from January 2000 to December 2016, yielding a sample size $n=204$.
However, the last $12$ observations have been reserved for forecasting comparison.
The data is freely available at INMET's website (\url{http://www.inmet.gov.br}).
 Figure \ref{f:rh} presents the time series plot (Figure \ref{f:rh-serie}) and the seasonal component in the data (Figure \ref{f:rh-sazonal}), sample autocorrelation (ACF) (Figure \ref{f:rh-fac}) and sample partial autocorrelation (PACF) functions (Figure \ref{f:rh-facp}).

\begin{figure}[p]
\begin{center}
\subfigure[Quantile residual]
{\label{f:residual}\includegraphics[width=0.4\textwidth] {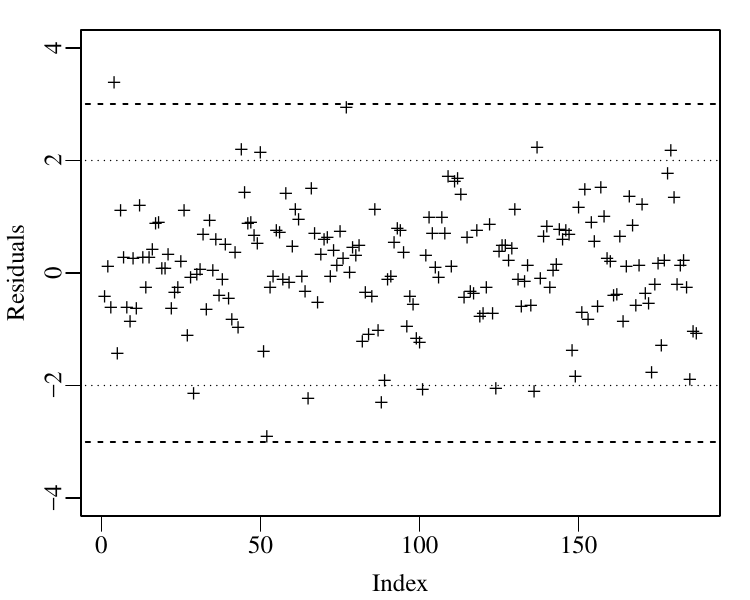}}
\subfigure[QQ-plot]
{\label{f:residual-qqplot}\includegraphics[width=0.4\textwidth] {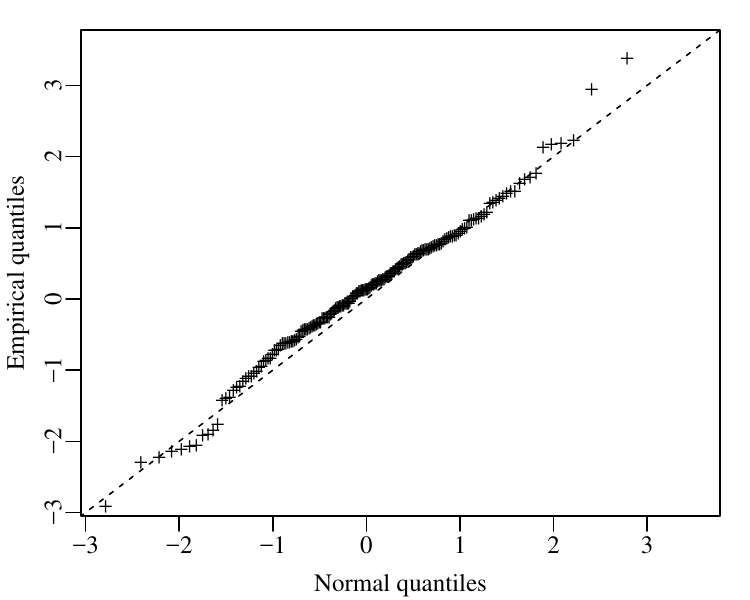}}
\subfigure[Residual ACF]
{\label{f:residual-fac}\includegraphics[width=0.4\textwidth] {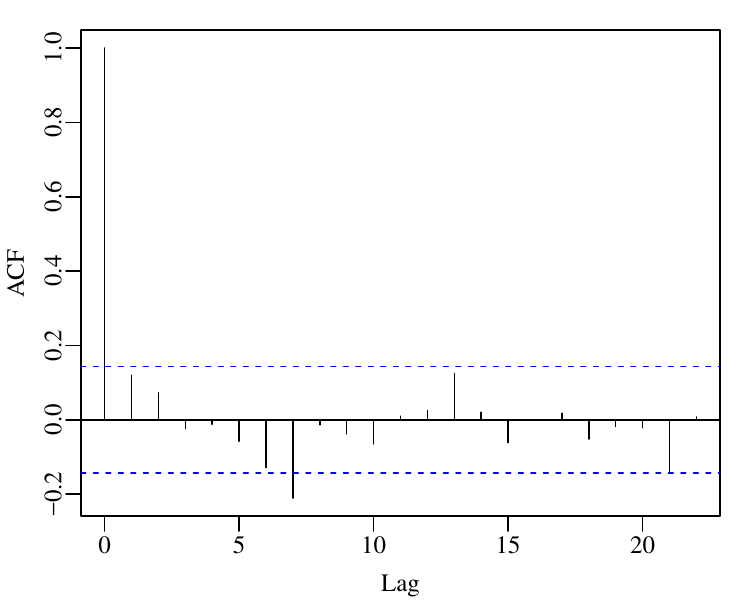}}
\subfigure[Residual PACF]
{\label{f:residual-facp}\includegraphics[width=0.4\textwidth] {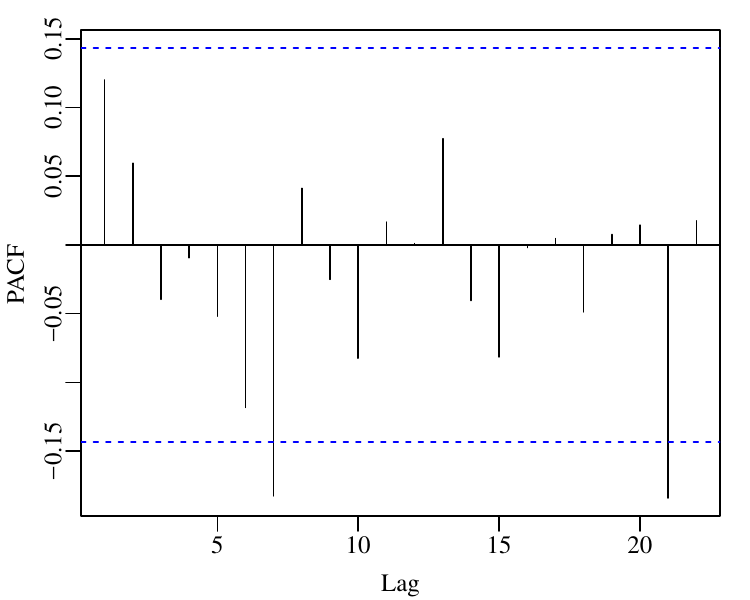}}
\caption{Diagnostic plots of the fitted KARMA model for relative humidity data based on the quantile residuals.}
\label{f:rh-diagnostico}
\end{center}

\begin{center}
{\includegraphics[width=0.5\textwidth]{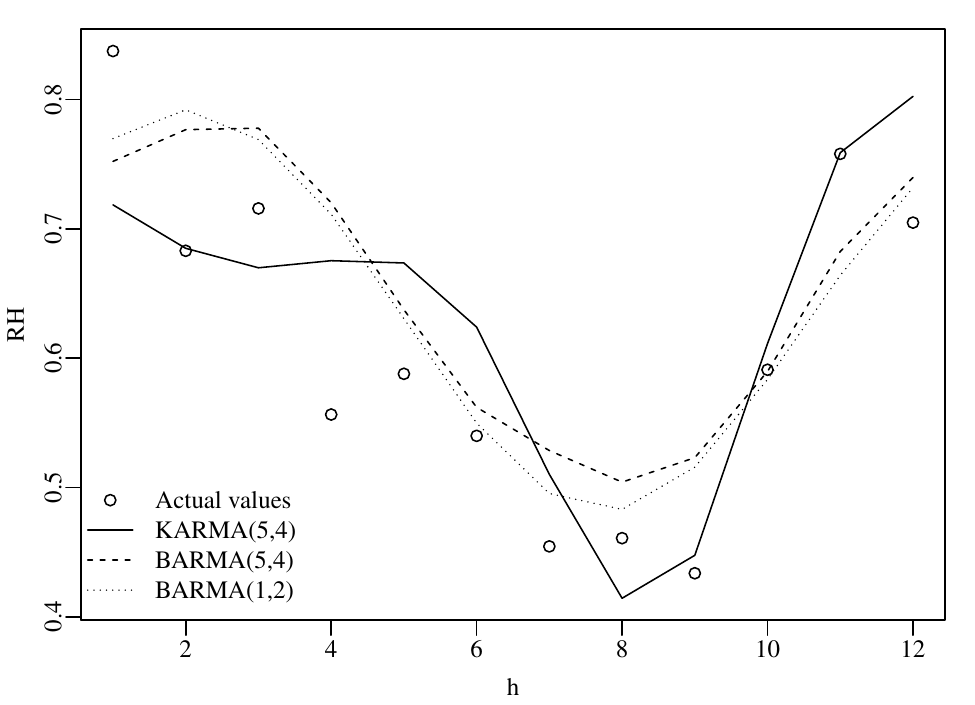}}
\caption{
Out-of-sample forecasting comparison.}\label{f:rh-forecast}
\end{center}
\end{figure}

From the Figures \ref{f:rh-serie} and \ref{f:rh-sazonal} we observe a clear seasonal component.
There are several ways to account for this monthly seasonal component.
We shall consider a simple harmonic regression approach \citep{Bloomfield2013},
by introducing the following covariates:
$x_t=\big(\sin(2\pi t/12),\cos(2\pi t/12)\big), \quad \mbox{for } \ t\in\{1,\dots,n\}$.
With the logit as link function,
using the three-stage iterative Box-Jenkins methodology \citep{Box2008} to select the fitted model,
we successfully modeled the data using a KARMA$(5,4)$ model with the covariates given above.
Table \ref{t:fittedkarma} brings the fitted KARMA model while Figure \ref{f:rh-diagnostico}
brings some residual diagnostic plots.
Figure \ref{f:residual} presents the residual plot against time. From this plot we observe no distinct pattern overtime and the typical white noise behavior for the residuals.
Figure \ref{f:residual-qqplot} shows the plot of the normal against the empirical quantiles. An approximately straight line, as seen in the plot, is an indication that the residuals are approximately normally distributed.
Finally the ACF and PACF shown in Figures \ref{f:residual-fac} and \ref{f:residual-facp}, respectively, can help to visually verify the residual white noise hypothesis, which was also test through the Ljung-Box test shown in Table \ref{t:fittedkarma}. All plots and tests indicate that the fitted model can be safely used for out-of-sample forecasting.

\begin{table}[t]
\caption{Fitted KARMA model for relative humidity data.} \label{t:fittedkarma}
\centering
\begin{tabular}{lcccc}
\hline
Parameter &	Estimate	&	Std. Error	&	$z$ stat.	&	${\rm Pr}(>|z|)$	\\
\hline
$\alpha	$ & $	0.8322	$ & $	0.1670	$ & $	4.9832	$ & $	0.0000	$ \\
$\phi_1	$ & $	1.2100	$ & $	0.1242	$ & $	9.7410	$ & $	0.0000	$ \\
$\phi_2	$ & $	-2.0237	$ & $	0.1196	$ & $	16.9168	$ & $	0.0000	$ \\
$\phi_3	$ & $	1.2470	$ & $	0.1600	$ & $	7.7932	$ & $	0.0000	$ \\
$\phi_4	$ & $	-1.0480	$ & $	0.1093	$ & $	9.5841	$ & $	0.0000	$ \\
$\phi_5	$ & $	0.2150	$ & $	0.0840	$ & $	2.5603	$ & $	0.0105	$ \\
$\theta_1	$ & $	-1.1076	$ & $	0.1396	$ & $	7.9345	$ & $	0.0000	$ \\
$\theta_2	$ & $	1.8808	$ & $	0.1773	$ & $	10.6099	$ & $	0.0000	$ \\
$\theta_3	$ & $	-0.8707	$ & $	0.1759	$ & $	4.9500	$ & $	0.0000	$ \\
$\theta_4	$ & $	0.8122	$ & $	0.1428	$ & $	5.6866	$ & $	0.0000	$ \\
$\varphi	$ & $	10.8870	$ & $	0.0741	$ & $	146.8605	$ & $	0.0000	$ \\
$\beta_1	$ & $	0.4134	$ & $	0.0319	$ & $	12.9428	$ & $	0.0000	$ \\
$\beta_2	$ & $	0.4465	$ & $	0.0358	$ & $	12.4810	$ & $	0.0000	$ \\

\hline
\multicolumn{5}{c}{AIC=$-487.8898$}\\
\multicolumn{5}{c}{Ljung-Box ($\text{lag}=20$): $Q=22.5520$ ($p$-value=$0.3113$)}\\
\hline
\end{tabular}
\end{table}

The
out-of-sample forecast of the adjusted KARMA model is presented in Figure \ref{f:rh-forecast}.
We observe that
the forecast was able to capture the distinctive seasonal pattern present in the actual data.
Figure \ref{f:rh-forecast} also shows
the  forecast values for the fitted $\beta$ARMA(5,4),
with the same order of the best KARMA model,
and
the $\beta$ARMA(2,1) which was the best $\beta$ARMA model.
In order to have a better comparison
we present some goodness-of-fit measures.
The
mean square error (MSE)
and
mean absolute percentage error (MAPE)
between the actual data ($y_{n+h}$) and
out-of-sample predicted ($\widehat\mu_{n+h}$) values,
for $h=1,\ldots,12$,
of the fitted models are presented in Table~\ref{t:comparison}.
We note that the proposed model
outperforms the $\beta$ARMA model in both measures.

\begin{table}[t]
\caption{Forecasting performance comparison among different models.} \label{t:comparison}
\centering
\begin{tabular}{lccc}
\hline
&	KARMA$(5,4)$	&	$\beta$ARMA$(5,4)$	&	$\beta$ARMA$(2,1)$	\\
\hline
MSE		&	$0.0050$ & $0.0060$ & $0.0053$\\
MAPE	&	$0.0961$ & $0.1133$ & $0.0989$\\
\hline
\end{tabular}
\end{table}

\section{Conclusions}\label{s:conclusion}

In this work we introduced a new class of dynamic regression models for double bounded time series.
More specifically, in the proposed KARMA$(p,q)$ models,
the conditional median of the Kumaraswamy distributed variable is assumed to follow a dynamic model involving covariates, an ARMA structure, unknown parameters and a link function.
Inference regarding KARMA model parameters is discussed and a conditional maximum likelihood approach is fully developed. In particular, closed expression for the score vector and the conditional Fisher information matrix are obtained. The conditional maximum likelihood approach is shown to produce consistent and asymptotically normal estimates. Based on the asymptotic results, the construction of confidence intervals and hypothesis testing is discussed. Diagnostic analysis and forecasting tools
are also discussed.
To assess finite sample performance of the CMLE in the KARMA framework, a Monte Carlo simulation study is performed. The simulation study showed that the CMLE performs very well even for small sample sizes. To exemplify its usefulness, an application of the KARMA model to monthly relative humidity data from Brasilia, the Brazilian capital city, is presented and discussed.

\section*{An {\tt R} implementation of the KARMA model}

An implementation in {\tt R} language \citep{R2017} to fit the KARMA model is available at \url{http://www.ufsm.br/bayer/karma.zip}.

\section*{Acknowledgements}

The authors acknowledge financial support from FAPERGS and CNPq, Brazil. 
The authors would also like to thank Professor Tarciana Liberal Pereira (UFPB, Brazil) for fruitful discussions and two anonymous referees for their valuable comments and suggestions which helped improving the quality of the first version of the paper. 
The authors would like to thank Professors Denise Botter and M\^onica Sandoval (IME/USP, Brazil)  for identifying a mistake in Lemma 2 of the original published article. 
This current version contains the correct results.

\section*{Appendix}\label{s:appendix-a}

In this appendix we present some technical lemmas and provide an outline of the proof of Theorem \ref{t:an}.

\appendix

\section{Lemma 1}\label{L:lema1}
%\begin{lemma}\label{L:lema1}
Let $Y_t$ be a random variable whose distribution given $\F_{t-1}$ is $K(\mu_t,\varphi)$. Then
\begin{align*}
\E\Big[\log(1-Y_t^\varphi)|\F_{t-1}\Big]=-\frac1{\delta_t}.
\end{align*}
Furthermore, if $\frac{\partial \ell_t}{\partial \mu_t}$ is given by \eqref{e:dldmu}, then $\E\Big(\frac{\partial \ell_t}{\partial \mu_t}|\F_{t-1}\Big)=0$.
%\end{lemma}

\noindent\emph{\textbf{Proof:}} Let $f_{\mu_t}$ be the density of $Y_t$, we have
\[\E\Big[\log(1-Y_t^\varphi)|\F_{t-1}\Big]=\int_0^1 \log(1-y^\varphi)f_{\mu_t}(y)dy.\]
Expanding $\log(1-y^\varphi)$ into a power series around 0, we obtain
\[\log(1-y^\varphi)=-\sum_{k=1}^\infty\frac{(y^\varphi)^k}{k}=-\sum_{k=1}^\infty \frac{y^{k\varphi}}{k},\]
from which it follows that
\begin{align*}
\int_0^1 \log(1-y^\varphi)f_{\mu_t}(y)dy = -\int_0^1\sum_{k=1}^\infty \frac{y^{k\varphi}}{k}f_{\mu_t}(y)dy=-\sum_{k=1}^\infty \frac{1}{k}\int_0^1y^{k\varphi}f_{\mu_t}(y)dy = -\sum_{k=1}^\infty \frac{1}{k}\E(Y_t^{k\varphi}|\F_{t-1}).
\end{align*}
From Proposition 3.2 in \cite{mitnik2013new}, it follows that, conditionally to $\F_{t-1}$, $Y_t^{k\varphi}\sim K(k,\delta_t )$ and hence
\begin{align*}
\E\Big[\log(1-Y_t^\varphi)\big|\F_{t-1}\Big]&=-\delta_t \sum_{k=1}^\infty \frac{1}{k}\B\left(k+1,\delta_t \right) .
%&=\frac{\log (2)}{\log(1-{\mu}_t^\varphi)}\sum_{k=1}^\infty \frac{1}{k}\B\left(k+1,\delta_t \right)
\end{align*}
Since $\Gamma(k+1)=k\Gamma(k)$, for positive integer $k$, for $a>0$
\begin{align*}
\frac{1}{k}\B(k+1,a)=\frac{\Gamma(k+1)\Gamma(a)}{\Gamma(a+k+1)}=\frac{k!\Gamma(a)}{k\Gamma(a)\prod_{i=0}^k(a+i)}=\frac{(k-1)!}{\prod_{i=0}^k(a+i)}=\frac1{a(a+1)}\binom{a+k}{a+1}^{-1}.
\end{align*}
Hence
\begin{align}\label{expect}
\E\Big[\log(1-Y_t^\varphi)\big|\F_{t-1}\Big]=-\frac1{\delta_t+1}\sum_{k=1}^\infty\binom{\delta_t+k}{\delta_t+1}^{-1}=-\frac1{\delta_t+1}\bigg[1+\frac1{\delta_t}\bigg]=-\frac1{\delta_t},
\end{align}
which is the desired result. The last assertion is a consequence of \eqref{expect} and \eqref{e:dldmu}.\hfill\qed

\section{Lemma 2}\label{L:lema2}
%\begin{lemma}\label{L:lema2}
If $Y_t$ is a random variable for which $Y_t|\F_{t-1}\sim K(\mu_t,\varphi)$,
then (for $\delta_t \not\in \{1,2\}$)
\[ \E\bigg(\frac{Y_t^{\varphi}\log(Y_t)}{1-Y_t^{\varphi}}\bigg|\F_{t-1}\bigg)= \frac{1-\psi(\delta_t+1)-\kappa}{(\delta_t-1)\varphi} \]
and
\[\E\bigg(\frac{Y_t^{\varphi}\log(Y_t)^2}{(1-Y_t^{\varphi})^2}\bigg|\F_{t-1}\bigg)=\frac{\delta_t\left(\psi(\delta_t)\big[\psi(\delta_t)+2(\kappa-1)\big]-\psi'(\delta_t)+k_0 \right)}{(\delta_t-2)(\delta_t-1)\varphi^2},\]
where $\psi:(0,\infty)\rightarrow\R$ is the digamma function defined as $\psi(z)=\frac{d}{dz}\log\big(\Gamma(z)\big)$, $\psi'(z)=\frac{d}{dz}\psi(z)$ is the trigamma function, $\kappa=0.5772156649\dots$  is the Euler-Mascheroni constant \citep{gradshteyn2007} and $k_0=\pi^2/6+\kappa^2 -2 \kappa$.
%\end{lemma}

\noindent\emph{\textbf{Proof:}} We have
\[\E\bigg(\frac{Y_t^{\varphi}\log(Y_t)}{1-Y_t^{\varphi}}\bigg|\F_{t-1}\bigg)=\varphi\delta_t\int_0^1y^{2\varphi-1}\log(y)(1-y^\varphi)^{\delta_t-2}dy.\]
Upon expanding $(1-y^\varphi)^{\delta_t-2}$ into its binomial series, we have $(1-y^\varphi)^{\delta_t-2}=\displaystyle{\sum_{k=0}^{\infty}}(-y^{\varphi})^k\binom{\delta_t-2}{k}$ and
\begin{align*}
\E\bigg(\frac{Y_t^{\varphi}\log(Y_t)}{1-Y_t^{\varphi}}\bigg|\F_{t-1}\bigg)&=\varphi\delta_t\sum_{k=0}^\infty(-1)^k\binom{\delta_t-2}{k}\int_0^1y^{\varphi(k+2)-1}\log(y)dy\\
&=\varphi\delta_t\sum_{k=0}^\infty(-1)^k\binom{\delta_t-2}{k}\bigg[-\frac{1}{\varphi^2(k+2)^2}\bigg]=-\frac{\delta_t}{\varphi}\sum_{k=0}^\infty\frac{(-1)^k}{(k+2)^2}\binom{\delta_t-2}{k}.
\end{align*}
To evaluate the series above, we change the index to $i=k+2$ and rewrite
\begin{align*}
\sum_{k=0}^\infty\frac{(-1)^k}{(k+2)^2}\binom{\delta_t-2}{k}&=\sum_{i=2}^\infty\frac{(-1)^{i-2}}{i}\bigg[\frac{1}{i}\binom{\delta_t-2}{i-2}\bigg]
=\sum_{k=1}^\infty\frac{(-1)^{i-2}}{i}\bigg[\frac{(i-1)}{\delta_t(\delta_t-1)}\binom{\delta_t}{i}\bigg]\\
&=\frac{1}{\delta_t(\delta_t-1)}\bigg[\sum_{i=1}^\infty(-1)^i\binom{\delta_t}{i}-\sum_{i=1}^\infty\frac{(-1)^i}{i}\binom{\delta_t}{i}\bigg].
\end{align*}

The result now follows by the Newton's series for the digamma function \citep[formula 8.363.8 in][with $n=0$]{gradshteyn2007}, that is,
\[\psi(s+1)+\kappa=-\sum_{k=1}^\infty\frac{(-1)^k}{k}\binom{s}{k},\]
and the identity $\sum_{k=1}^\infty(-1)^k\binom{n}{k}=-1$. Similar technique yields
\begin{align*}
\E\bigg(\frac{Y_t^{\varphi}\log(Y_t)^2}{(1-Y_t^{\varphi})^2}\bigg|\F_{t-1}\bigg)&=\varphi\delta_t\sum_{k=0}^\infty(-1)^k\binom{\delta_t-3}{k}\int_0^1y^{\varphi(k+2)-1}\log(y)^2dy\\
&=\varphi\delta_t\sum_{k=0}^\infty(-1)^k\binom{\delta_t-3}{k}\bigg[\frac{2}{\varphi^3(k+2)^3}\bigg]=\frac{2\delta_t}{\varphi^2}\sum_{k=0}^\infty\frac{(-1)^k}{(k+2)^3}\binom{\delta_t-3}{k}.
\end{align*}
The result now follows by similar argument as the previous case and from the Newton's series for the digamma and $\psi'(x)$ functions \citep[formula 8.363.8 in][with $n=0,1$]{gradshteyn2007}, that is,$\psi'(x)=\sum_{k=0}^\infty\frac{1}{(x+k)^2}$.\hfill\qed

\section{Proof of Theorem \ref{t:an}}\label{a:proofteo}
\noindent\emph{\textbf{Proof:}}
 In order to obtain the results, we only need to check that assumptions 2.1-2.5 from \cite{Andersen1970} are fulfilled.  Assumption 2.1 follows from Section \ref{s:score}.
 Assumption 2.2 follow from standard results for ARMA models with covariates \citep{Hannan1973}.
 To show that Assumption 2.3 holds, observe that, for small $\delta$ in a neighborhood of 0, the argument for the variance can be written as (recall that conditionally on the past, $y_t$'s are independent)
\begin{align*}
\sum_{t=m}^n &\ell_t(\mu_t,\varphi+\delta)-\ell_t(\mu_t,\varphi)
= \sum_{t=m}^n \delta \log(y_t)+J_{\varphi+\delta}\log(1-y_t^{\varphi+\delta})-J_{\varphi}\log(1-y_t^{\varphi})+ C_t
\end{align*}
where $J_{x}=\log\Big(\log\big(0.5/\log(1-\mu_t^x)\big)\Big)$ and $C_t$ are (non-random) real constants. The terms $\log(y_t)$, $\log(1-y_t^{\varphi+\delta})$ and $\log(1-y_t^{\varphi})$ can be shown to be continuous functions
of their arguments so that the result follows \citep[see][]{Andersen1970}.  Assumption 2.3 are satisfied by the definition of the KARMA model and the results on Section \ref{s:estimation}. Assumption 2.4 is a consequence of Lemma 1 in  \ref{L:lema1} and the final condition follows from the assumptions on the design (covariate) matrix and Section \ref{s:inf}.\hfill\qed

\section*{References}

%\bibliographystyle{elsarticle-harv}
%\bibliography{karma}

%\end{linenomath}
%\end{linenumbers}
\end{document}